\documentclass{article}

\usepackage{arxiv}
\usepackage{graphicx}
\usepackage[utf8]{inputenc} % allow utf-8 input
\usepackage{hyperref}       % hyperlinks
\usepackage{url}            % simple URL typesetting
\usepackage{booktabs}       % professional-quality tables
\usepackage{amsfonts}       % blackboard math symbols
\usepackage{nicefrac}       % compact symbols for 1/2, etc.
\usepackage{microtype}      % microtypography
\usepackage{xcolor}

\title{Extrinsic Voltage Control of Carrier Lifetime in Polycrystalline $PbSe$ Mid-wave IR Photo Detectors for Increased Detectivity}

\author{
Samiran Ganguly, Avik W. Ghosh \\
Charles L. Brown Dept. of Electrical and Computer Engineering, University of Virginia, Charlottesville, VA\\
Tang Xin, Philippe Guyot-Sionnest\\
Dept. of Chemistry and Dept. of Physics, University of Chicago, Chicago, IL\\
Sung-Shik Yoo \\
Northrop Grumman Systems Corp., Rolling Meadows, IL\\
}

\begin{document}
\maketitle

\begin{abstract}
Polycrystalline $PbSe$ for mid-wave IR (MWIR) photodetector is an attractive material option due to high operating/ambient temperature operation and relatively easy and cheap fabrication process, making it candidate for low-power and small footprint applications such as internet-of-thing (IoT) sensors and deployment on mobile platforms due to reduced/removed active cooling requirements. However, there are many material challenges that reduce the detectivity of these detectors. In this work, we demonstrate that it is possible to improve upon this metric by externally modulating the lifetime of conducting carriers by application of a back-gate voltage that can control the recombination rate of generated carrier. We first describe the physics of $PbSe$ detectors, the mechanisms underlying carrier transport, and long observed lifetimes of conducting carriers. We then discuss the voltage control of these inverted channels using a back-plane gate resulting in modulation of the lifetime of these carriers. This voltage control represents and extrinsic ``knob'' through which it may be possible to open a pathway for design of high performance IR photodetectors, as shown in this work.
\end{abstract}

\section{Introduction}

Mid-wave Infrared (IR) atmospheric detection window ($3 - 8~\mu m$ of EM spectrum) has been useful for applications in astronomy  and earth sciences \cite{noauthor_wide-field_nodate,noauthor_instruments_nodate} as well as military and surveillance \cite{corsi_infrared_2012}. Integration of imaging sensors in commercial civilian applications such as self-driving automotives \cite{thakur_infrared_2017} may expand the range of application of such sensors many-fold. These new classes of sensors will need to have high resolution and performance without active cooling to reduce energy and production cost, and increase compactness and integrability with other systems.

Therefore, there is a need for novel designs of photo-detectors that can meet such targets. $PbSe$ is a century old material \cite{rogalski_infrared_2002,rogalski_history_2012} that only has a niche market in mid-wave IR, compared to HgCdTe detectors, even though its cost of production is low and is easily integrable with a $Si$ based read-out integrated circuit (ROIC) \cite{roelkens_silicon-based_2013}. This is due to relatively lower performance and higher device-to-device variability in a large focal plane array. Better controlled fabrication protocols helps in mitigating these issues to an extent, however, it is expedient to explore extrinsic electrical control that a ROIC can provision to control these factors.

In these detector the central working principle is modification of the detector film resistivity in presence of incident photons through generation of new charge carriers in the film \cite{ganguly_multiscale_2019} and the ``signal'' depends on the difference between the background doping (dictates dark current) and the resulting carrier concentration (dictates lit current). The number of photo-generated carrier above the background doping can be improved in a few ways. One approach is through plasmonic enhancement of absorption (see e.g.\cite{wittenberg_surface_2017,grayer_embedded_2019,rabiee-golgir_ultra-thin_2019}) which can improve the quantity of photo-generated current. Another approach is through appropriate choice of blocking contact material \cite{ganguly_choice_2019} that can recycle the trapped electrons over many holes which enhances the total number of photo-generated carriers seen by the ROIC. In this work, we discuss another technique focusing purely on enhancing the lifetime $\tau$ of the photo-generated holes by reduction of their recombination in the film through a back-gate voltage. 

In this paper we first briefly discuss the detector fabrication and characterization setup. These were performed at the Northrop Grumman Corp. and University of Chicago. Then we discuss the physics of these detectors and the connection between the carrier lifetime with detector performance. We also provide a benchmark against our analytical model for the device against the fabricated device's electrical characteristics. Then we build a semi-analytical model for lifetime vs. the back-gate voltage and benchmark against the experimental device. In doing so we also build a semi-empirical model of carrier lifetime vs. temperature in these detectors. Finally we provide analysis of performance improvement seen by the experimental detector due to back-gate voltage and design optimizations that can translate the small signature of performance improvement in the experimental device into substantial increase in well-designed detectors.

\begin{figure}[htbp]
\centering
\includegraphics[width=3in]{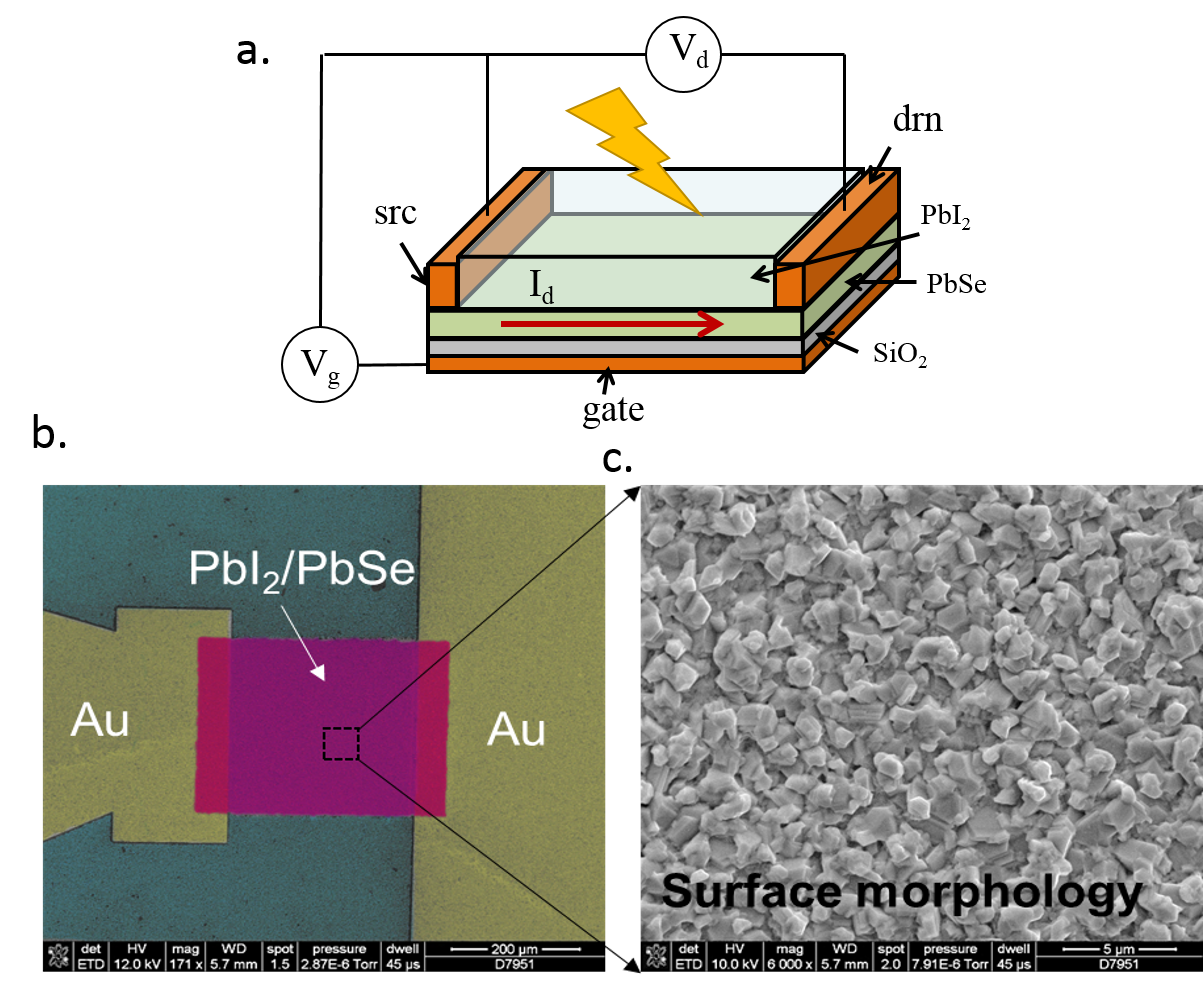}
\caption{a. Schematic of the back-plane gated $PbSe$ photodetector. The gate voltage is applied through the $Si|SiO_2$ substrate, while source and drain contacts carry dark and photo-generated current to a ROIC. b. Plan view image of the actual patterned detector with $Au$ leads. c. Surface micrograph of the film showing its polycrystalline nature. }
\label{fig:1}
\end{figure}

\section{Detector Fabrication and Characterization}
\label{sec:expt}
$PbSe$ films were fabricated using the standard liquid epitaxy method commonly called the chemical bath deposition, where the organic precursors containing $Pb$ and $Se$ are allowed to chemically combine and precipitate on the substrate. This film is taken through two further steps of annealing, once in an Oxygen environment, and then in an Iodine environment. These steps increase the film's light sensitivity significantly and Hall measurements find the film to have undergone carrier inversion and significant reduction in mobility. The top half of the film also converts to $PbI_2$ which act as a natural capping layer, but also enables high photo-sensitivity as it has been observed and has been part of $PbSe$ detector fabrication for nearly four decades \cite{johnson_lead_1983}. 

The sensitized $PbSe$ films on $SiO_2|Si$ substrate were then patterned into squares to define the active sensing area (fig. \ref{fig:1}b). Photoresist AZ703 was used to define and cover the sensing area by UV lithography, followed wet chemical etching in hydrogen peroxide bath for $\sim 10~min$. After patterning, the residual photoresists were removed by rinsing in acetone and isopropanol. Electrodes were defined by photolithography and deposition of $10~nm$  $Ti$ and $50~nm$ $Au$ by e-beam evaporation. Lift-off process was conducted in acetone bath for $10~min$, dried by nitrogen blow. 

Blackbody source with controlled temperature was used as the light source in current-voltage $(I_d-V_d)$ and field effect transistors $(I_d-V_g)$ measurements. The photocurrent was first amplified with a trans-impedance amplifier (DLPCA-200, Femto) and then amplified by a voltage amplifier. The applied bias voltages can be tuned by the built-in bias of the transimpedance amplifier. The amplified photocurrents were digitalized and recorded by a data acquisition card. 

\section{Detector Physics, Performance Metrics, and Performance Enhancement}
\label{sec:detector}

Previously we \cite{ganguly_multiscale_2019} and others \cite{qiu_study_2013,zhao_understanding_2014} have discussed the critical role that is played by the interfacial depletion field between $PbSe$ and $PbI_2$ in enhancing the lifetime of the conducting holes in the photo-sensitive $PbSe$ detector films. These depletion film formed at the interface of the $PbSe|PbI_2$ separate the photo-generated excitons in to electrons and holes. The electrons are trapped within the grain boundaries rich in $PbO^{++}$ species, while holes are available for conduction. In these polycrystalline films the transport of the conducting carriers, in this case the holes, can be modeled as the hopping from one conducting site to the other over the barriers. We also derived an expression for the current density-applied voltage relationship through these films in the photo-conductive mode given as:

\begin{equation}
|J| = 2q(p_0+G\tau/t_{film})\mu_p\exp(\frac{-q\phi_b}{k_BT})\sinh(\frac{qV_{d}}{2N_bk_BT})\frac{\phi_b}{t_b}
\label{eq:lightcurrentdensity-scalar}
\end{equation}

In this equation $J,p_0, G,\tau,t_{film},\phi_b,V_{d},N_b,\mu_p,t_b,T$ respectively are the current density magnitude, intrinsic carrier concentration, photo-carrier generation rate, hole lifetime, film thickness, average inter-crystalline barrier height, applied voltage bias, total number of barriers presented by the film in the transport direction, average barrier thickness, hole mobility, and temperature of the detector. $q,k_B$ are the usual physical constants they represent. The parameters $N_b, t_b$ can be obtained from imaging techniques such as scanning electron microscopy where average crystallite dimensions and inter-crystallite distances can be estimated. 

It can be seen from eq. \ref{eq:lightcurrentdensity-scalar} that the photo-generated carriers increase the conductivity of the film and therefore it presents a lower resistance, that can be measured by the ROIC by measuring the $RC$ delay of a  network formed using the detector and a controlled capacitor. The generation rate $G$ depends on the quantum efficiency, i.e. the efficiency of conversion of light into carriers and the light absorption. 

Two important metrics for photo-detectors are responsivity and specific detectivity, which measure photo-signal to electrical-signal transduction and signal-to-noise ratio and are given by: 

\begin{eqnarray}
\mathcal{R} &=& \frac{I_{photo}}{P_{optical}}  \propto \frac{G\mu\tau}{P_{optical}}\\
\mathcal{D^*} & \propto & \frac{I_{photo}}{I_{noise}} \propto \frac{G\mu\tau}{I_{noise}}
\label{eq:perf}
\end{eqnarray}

Where $ I_{photo},P_{optical},G, I_{noise}$ stand for generated photo-current (given by: $I_{photo} = I_{lit}-I_{dark}$), incident optical power, carrier generation rate, and noise current in the detector. It can be seen that all these three metrics depend on the lifetime of the transporting carriers. Therefore, increasing the carrier lifetime in critical to obtain higher performance from such detectors.

In ref. \cite{ganguly_multiscale_2019} we developed a simple model of the carrier lifetime ($\tau$) in terms of partial derivative of the total recombination rate $R$ with electron and hole concentrations ($n,p$), and the fraction of depleted $PbSe$ layer $r_{surf}$:

\begin{equation}
\tau^{-1} \approx \frac{\partial R}{\partial \Delta p} + \left(1-r_{surf}\right) \frac{\partial R}{\partial \Delta n}
\label{eq:lifetime}
\end{equation}

where the total recombination rate is sum of the partial rates from the recombination phenomena in the film, viz. direct/radiative ($PbSe$ is a direct bandgap material), Shockley-Reed-Hall (SRH), and Auger. The surface depletion from the $PbSe|PbI_2$ interface and its effect on the lifetime through suppression of non-conducting carrier, which is electrons in this, and hence the factor $r_{surf}$ appears with the partial derivative w.r.t. electrons. We argue that by building a back gate into the detector structure, we can electrostatically control $r_{surf}$ and therefore control the lifetime of the detector which results in increase in the metrics of performance: responsivity and detectivity.

\begin{figure}[htbp]
\centering
\includegraphics[width=\linewidth]{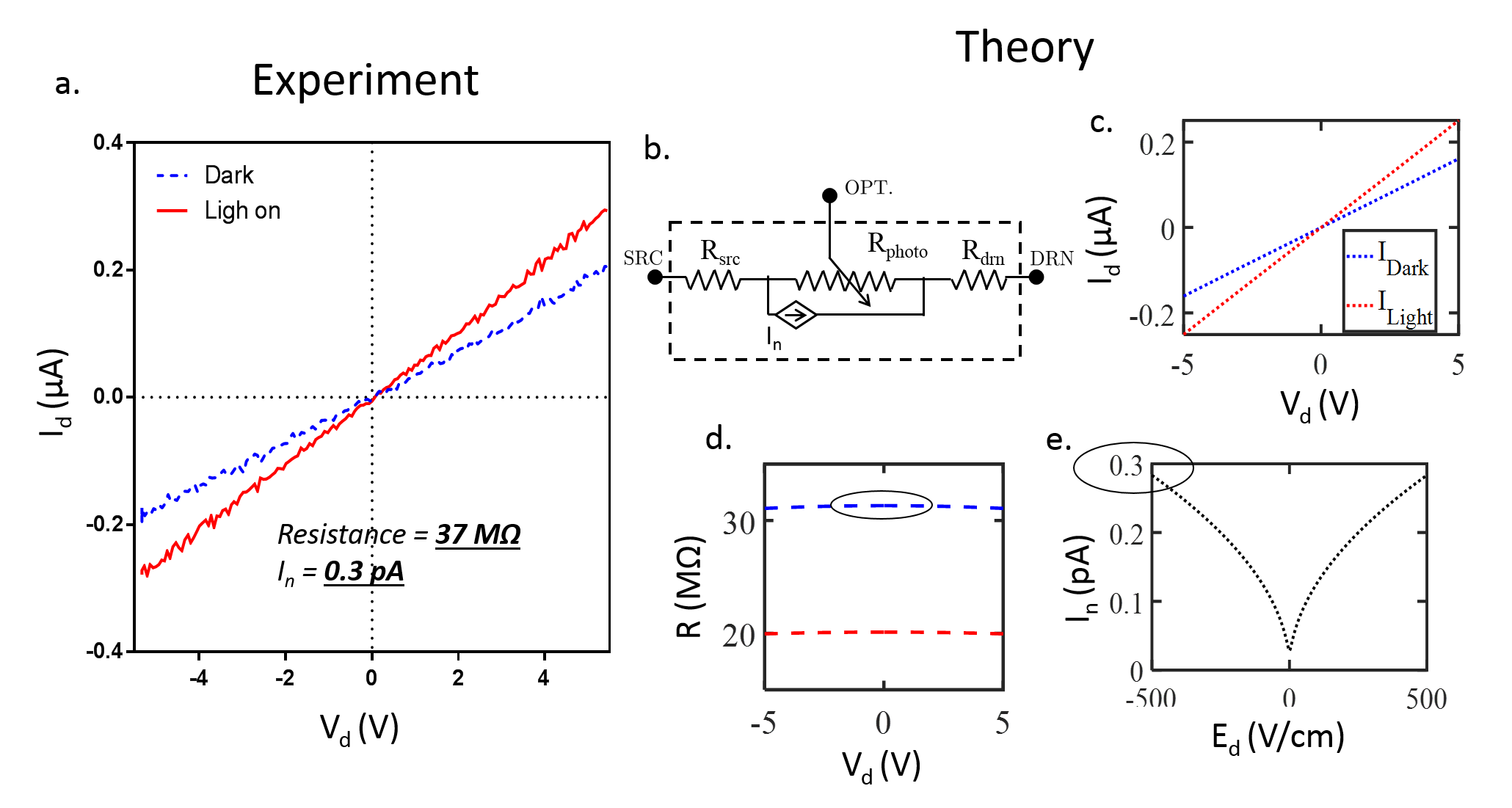}
\caption{Benchmarking of detector characteristics a. Measured $I_d-V_d$ characteristics, $R$, $I_{noise}$ b. Circuit model for the detector as a parallel combination of controlled resistor and noise current source, with source and drain resistances also accounted for. c. Modeled $I_d-V_d$ characteristics c. Modeled $R-V_d$ characteristics d. Modeled $I_{noise} - E_d$ characteristics. }
\label{fig:3}
\end{figure}

In fig. \ref{fig:3}a we show  the experimental electrical characteristics and noise in the fabricated detectors, and the analytical model (eq. \ref{eq:lightcurrentdensity-scalar}) as a circuit model also accounting for the contact resistances (fig. \ref{fig:3}b) . It can be seen that we can very closely match the $I-V$ characteristics (fig. \ref{fig:3}c), film resistance (fig. \ref{fig:3}d) as well as the noise magnitude (fig. \ref{fig:3}e) using the model (parameters listed in tab. \ref{tab:1}) and is close to those used in ref. \cite{ganguly_multiscale_2019}.

\begin{table}%
\begin{tabular}{|l|l||l|l|}
\hline 
\textbf{Parameter} & \textbf{Value} & \textbf{Parameter} & \textbf{Value} \\
\hline 
Area ($L\times W$) & $100 \times 100\mu m^2$ & $t_{PbSe}$ & $0.5 \mu m$ \\
$t_{PbI_2}$ & $1\mu m$ & $t_{ch;0}$ & $18 nm$\\
$t_b$ & $20nm$ & $w_b$ & $0.2\mu m$ \\
$t_{ox}$ & $0.3\mu m$ & $T$ & $230-295K$\\
$p_0$ & $1.2\times 10^{15}/ cm^{3}$ & $E_g$ & $0.29 eV$\\
$E_b$ & $265meV$ & $\mu(295K)$ & $0.1 cm^2/ V s$ \\
$G$ & $6.68\times10^{18} /cm^{2}s^{1} $ & $P_{opt}$ & $23 \mu W /cm^{2}$ \\
$\rho_{SRH} (250K)$ & $170 \mu s$ & $\rho_{Aug} (250K)$ & $155 \mu s$\\
$\kappa_{SRH}$ & $9$ & $\kappa_{Aug}$ & $4.1$\\
\hline
\end{tabular}
\caption{Parameter list used in the back-gated PbSe detector model.}
\label{tab:1}
\end{table}

\section{Lifetime Modulation using Back-gating}
\label{sec:backgate}

%The $PbSe$ photodetector is a planar structure with the top surface being a $PBI_2$ capping and passivation layer transparent to mid-wave IR wavelengths. The bottom layer is therefore open to electrical control. We can grow the detector on top of a heavily doped $Si$ layer meant to work as the back-gate terminal and field oxide composed of $SiO_2$. This can then be attached to a ROIC capable of providing a gate bias to the detector. The reference voltage point can be provided by the source terminal built on the top side of the $PbSe$ film. This gives rise to slanted electric fields that should still be effective in modulating the depletion volume in the detector, but it can be improved if a transparent grounded top contact can be grown to provide more uniform gate control over the detector.

\begin{figure}[htbp]
\centering
\includegraphics[width=\linewidth]{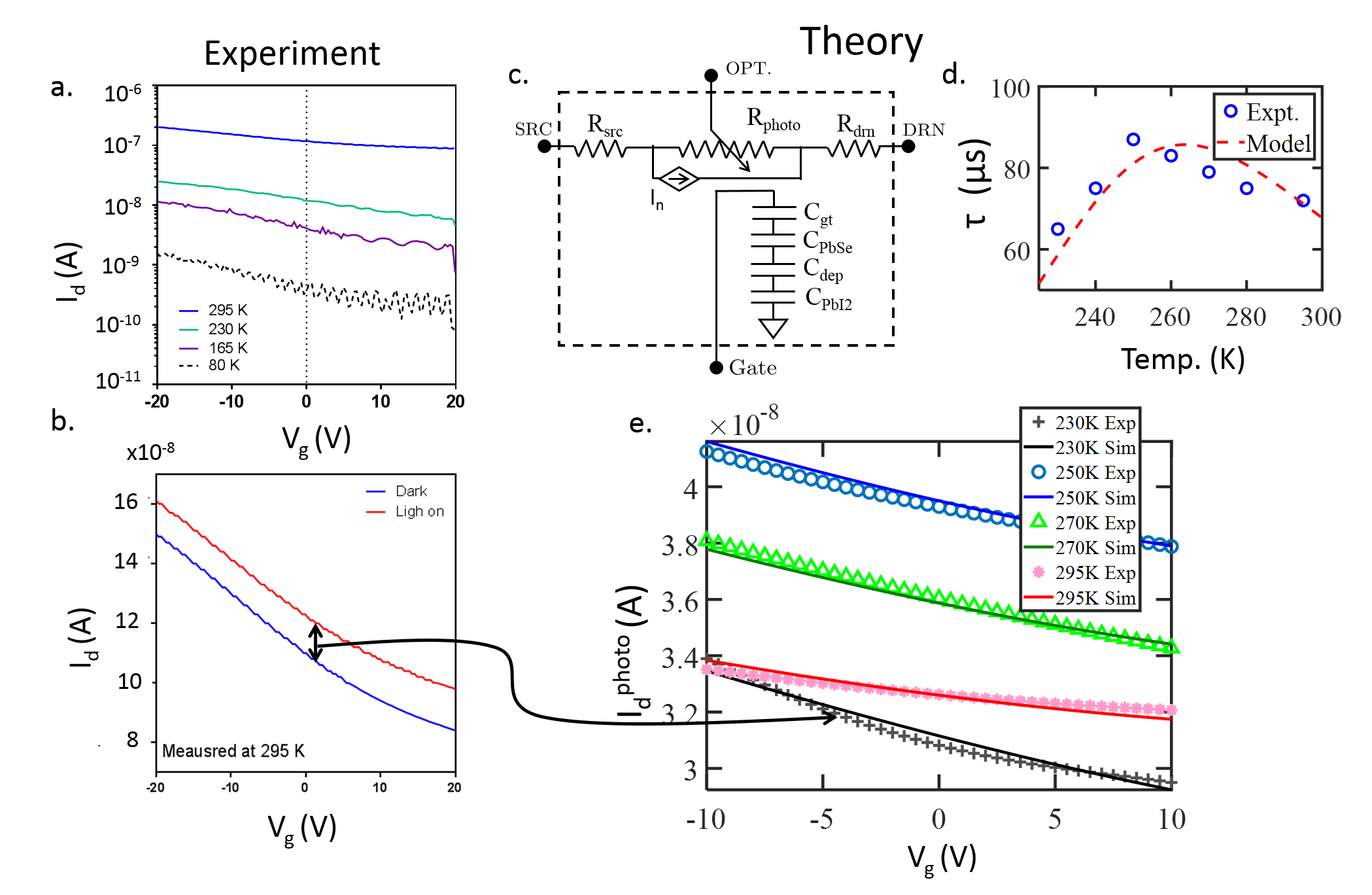}
\caption{Benchmarking of back-gating effect a. Measured $I_d-V_g$ characteristics at different temperatures. b. Measured $I_d-V_g$ characteristics for lit and dark conditions at $295K$. c. Measured vs. Modeled $\tau - Temp.$ characteristics. d. Modeled $I_d^{photo}-V_g$ characteristics at different temperatures  }
\label{fig:4}
\end{figure}

We measure the modulation of dark and lit current as a function of gate voltage for a constant drain bias (fig. \ref{fig:4}a,b) and develop a numerical capacitor-divider based model (fig. \ref{fig:4}c) for the electrostatics of the detector stack, by considering a series network of the back-gate oxide, $PbSe$, depleted layer, and $PbI_2$ layer. We then incorporate the effect of the applied gate voltage on the ratio of the depletion layer thickness to the $PbSe$ layer thickness, assuming uniform depletion along the plane of the detector film. This ratio gives us the $r_{surf}$ parameter for the eq. \ref{eq:lifetime}, thereby yielding a semi-analytical formulation for the lifetime of the carriers.

However, the lifetime is also a function of temperature as at different temperature range different recombination mechanisms dominate. In particular we focus on multi-particle processes such as Shockley Reed Hall (SRH) and Auger processes which are mediated by trapping centers and phonons respectively. At low temperatures, the trapping centers become ``deeper'', as the thermal motion reduce which reduces the likelihood of a trapped particle to escape. Similarly at high temperatures the phonons increase in number (by Bose-Einstein statistics) which increase their likelihood to cause Auger recombination. From experimental measurements of lifetime we observe that it shows a non-monotonic behavior with temperature that can be explained by the heuristic presented above. We can therefore model this using the following phenomenological equation:

\begin{eqnarray}
\tau_{SRH} &=& \rho_{SRH} T^{\kappa_{SRH}}\\
\tau_{Aug} &=& \rho_{Aug} T^{-\kappa_{Aug}}\\
\frac{1}{\tau_0} &=& \frac{1}{\tau_{SRH}} + \frac{1}{\tau_{SRH}}
\label{eq:tau_T}
\end{eqnarray}

Where the $\rho_{SRH},\rho_{Aug},\kappa_{SRH},\kappa_{Aug}$ are fitting parameters to match the experimental trends, $\tau_{SRH},\tau_{Aug},\tau_0$ are lifetime contributions from SRH and Auger mechanisms as the total lifetime at $0$ gate-bias. We see a good fit (fig. \ref{fig:4}d) with the observed non-monotonic trend of the lifetime with temperature. Using the calculated $\tau_0$ from the above model, we can recast the eq. \ref{eq:lifetime} in the following form:

\begin{equation}
\tau = \frac{\tau_0}{1-[\frac{t_{ch}-t_{ch;0}}{t_{PbSe}-t_{ch;0}}]^{\gamma}}
\label{eq:tau_V}
\end{equation}

Where $t_{ch},t_{ch;0},t_{PbSe}$ are the depletion width at an applied gate-bias, built-in depletion width, and the $PbSe$ layer thickness respectively, $\gamma$ is a fitting exponent that ranges from $0.9$ to $1.2$ in the range of temperatures measured. The depletion layer thickness can be calculated using the electrostatic model as discussed above.

As per the eq. \ref{eq:lifetime}, the exponent $\gamma$ should ideally be $1$, but needs slight adjustment to account for simplifications built into the model, e.g. ignoring direct recombination, and the effect of lifetime due to modulation of hole density (considered to be 0) as holes are majority conducting-carriers in the depletion region and their concentration does not change significantly. Using \ref{eq:tau_V} we can capture the experimentally measured trends of lifetime on back-gate bias quite closely (fig. \ref{fig:4}e). This demonstrates the validity of the back-gating approach in improving the lifetime of the carriers extrinsically.

\section{Result: Detector Performance Enhancement}
\label{sec:perf}

As noted in eq. \ref{eq:perf} performance metric like $\mathcal D^*$, benefit from increased carrier lifetime, as they enhance the signal. While we did not measure the $\mathcal D^*$ itself directly, it is easy to estimate that it will be modulated by $6-20\%$ over the voltage range shown in fig. \ref{fig:4}e directly due to the lifetime modulation on the experimental detector, depending on the operating temperature ($295K-230K$). 

\begin{figure}[htbp]
\centering
\includegraphics[width=3in]{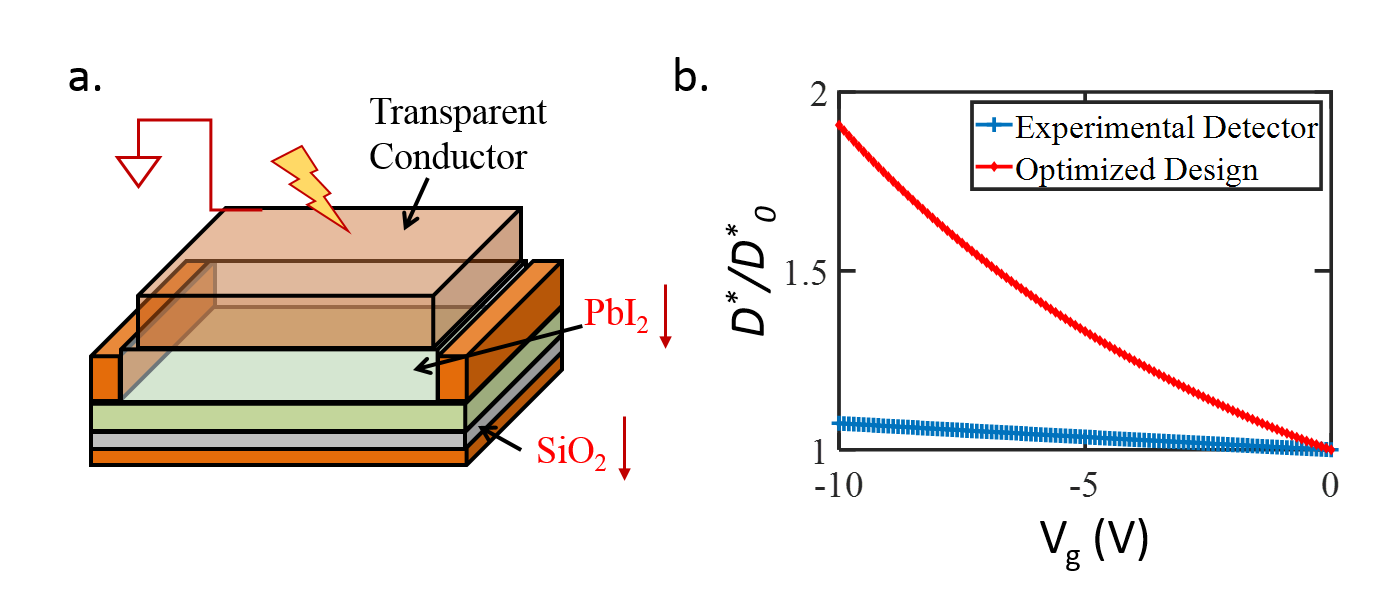}
\caption{Projection of back-gating effect on performance a. Optimized detector design, introduction of a transparent top ground-plane, thinning of gate oxide, and $PbI_2$ layers. b. Modeled $D^*-V_g$ characteristics for experimental detector to optimized detector design. }
\label{fig:5}
\end{figure}

In fig. \ref{fig:5}a we show the schematic of the optimized design. The oxide layer is thinned significantly to $10~nm$, the thickness of $PbI_2$ layer is halved ($1000~nm$ to $500~nm$), and a top surface full coverage ground plane assumed (can be fabricated from mid wave IR transparent contacts such as ITO or graphene). It can be seen from fig. \ref{fig:5}b that for the detector at $250K$, the improvement of $\mathcal D^*$ between zero back-gate bias to the high back-gate bias of $-10V$ is $6\%$ in experimental device while it is $90\%$ in the optimized detector. Other possible optimizations, which we do not perform here, include modulating the doping of the $PbSe$ film and optimizing the film thickness of $PbSe$ layer, though the film cannot be thinned significantly as this reduces the absorption of light in the detector which harms the signal. This demonstrates that extrinsic gate control of lifetime can be an effective means in improving the performance of $PbSe$ mid-wave IR photo-detectors.

\section*{Acknowledgments and Data Availability}

Research funded by DARPA/MTO under the WIRED program contract no. FA8650-16-C-7637. The authors thank Dr. Justin Grayer for useful discussions. The data that supports the findings of this study are available within the article.

% Bibliography
%\bibliographystyle{unsrt} 
%\bibliography{PbSephotodetector}

\end{document}